\documentclass[a4paper,
              ]{jacow}
\makeatletter%
	\ifboolexpr{bool{xetex}}
	 {\renewcommand{\Gin@extensions}{.pdf,%
	                    .png,.jpg,.bmp,.pict,.tif,.psd,.mac,.sga,.tga,.gif,%
	                    .eps,.ps,%
	                    }}{}
\makeatother

\ifboolexpr{bool{xetex} or bool{luatex}} 
 {}                                      
 {\usepackage[utf8]{inputenc}}           

\usepackage[USenglish]{babel}			 

\usepackage[final]{pdfpages}
\usepackage{multirow}
\usepackage{ragged2e}
\usepackage{hyperref}
\usepackage{fancyhdr}

\ifboolexpr{bool{jacowbiblatex}}%
 {%
  \addbibresource{jacow-test.bib}
  \addbibresource{biblatex-examples.bib}
 }{}
\begin{document}
\fancypagestyle{plain}{%
   \fancyhead[R]{\fbox{FERMILAB-CONF-21-684-AD}}
   \renewcommand{\headrulewidth}{0pt}
}

\title{Electron Cooling With Space-Charge Dominated Proton Beams at \NoCaseChange{IOTA}}
\author{N. Banerjee\thanks{nilanjan@uchicago.edu}, M.K. Bossard, J. Brandt, Y-K. Kim, The University of Chicago, Chicago, USA \\
		B. Cathey, S. Nagaitsev\thanks{also at the University of Chicago}, G. Stancari, Fermilab, Batavia, Illinois, USA}
		
\maketitle
\begin{abstract}
    We describe a new electron cooler being developed for 2.5 MeV protons at the Integrable Optics Test Accelerator (IOTA), which is a highly re-configurable storage ring at Fermilab. This system would enable the study of magnetized electron cooling in the presence of intense space-charge with transverse tune shifts approaching -0.5 as well as highly non-linear focusing optics in the IOTA ring. We present an overview of the design, simulations and hardware to be used for this project.
\end{abstract}

\section{Introduction}
The creation and stability of high-intensity hadron beams is very important to future projects such as heavy-ion facilities\cite{Spiller2020,Yang2013,Kekelidze2017}, Electron-Ion Colliders\cite{Willeke2021}, etc. Electron cooling provides a well-established method of attaining high equilibrium beam intensities and have been demonstrated for a wide range of ion energies from $\gamma \sim 1.00011$\cite{Bartmann2018} up to $\gamma \approx 9.5$\cite{Nagaitsev2006}. The maximum intensity of ion beams achieved through electron cooling is limited by the additional heating processes of Intra-Beam Scattering (IBS) and resonance-driven transverse heating due to space-charge tune shifts.\cite{Parkhomchuk2001} In practice, the transverse size of the ion beam decreases under the influence of electron cooling until the betatron tune shift reaches a maximum value of 0.1-0.2.\cite{Nagaitsev95,Steck2000} Studying the influence of ion space-charge forces on electron cooling at the high-intensity limit requires the development of a novel test platform and associated theoretical models.\\

The Integrable Optics Test Accelerator (IOTA) is a re-configurable 40~m storage ring built at Fermilab which acts as a test facility dedicated to research on intense beams including the areas of Non-linear Integrable Optics (NIO), beam cooling, space-charge, instabilities and more.\cite{Antipov2016,Valishev2021} It can circulate both electrons up to 150~MeV and protons with a kinetic energy of 2.5~MeV ($pc \approx 70$ MeV).  The proton beam energy is limited by the existing injector RFQ. In this contribution, we discuss the design of the electron cooler which we will operate with 2.5~MeV protons as a part of our electron-lens research program.\cite{Stancari2021} Besides enabling experiments on non-linear integrable optics by reducing energy spread and improving the lifetime of the beam by compensating for transverse emittance growth, the primary motivation of research with this cooler is to study the effect of space-charge forces in the regime of large transverse incoherent tune shift of $\Delta \nu_{x,y} \approx -0.5$ and compare with theoretical models. In addition, we are also planning experiments which uses electron cooling as a knob to study the interplay between space-charge and instabilities\cite{Burov2018} and also control the phase space distribution in order to facilitate the realization of NIO in the presence of space-charge forces.\\

In the next section, we detail the operation parameters of IOTA with protons and describe the electron cooler setup. Then we discuss a novel simulation model which includes electron cooling with transverse space-charge. In the last section, we summarize our results and present future plans.\\

\section{Electron Cooler Setup}

\begin{figure}
    \centering
    \includegraphics[width=\textwidth/2]{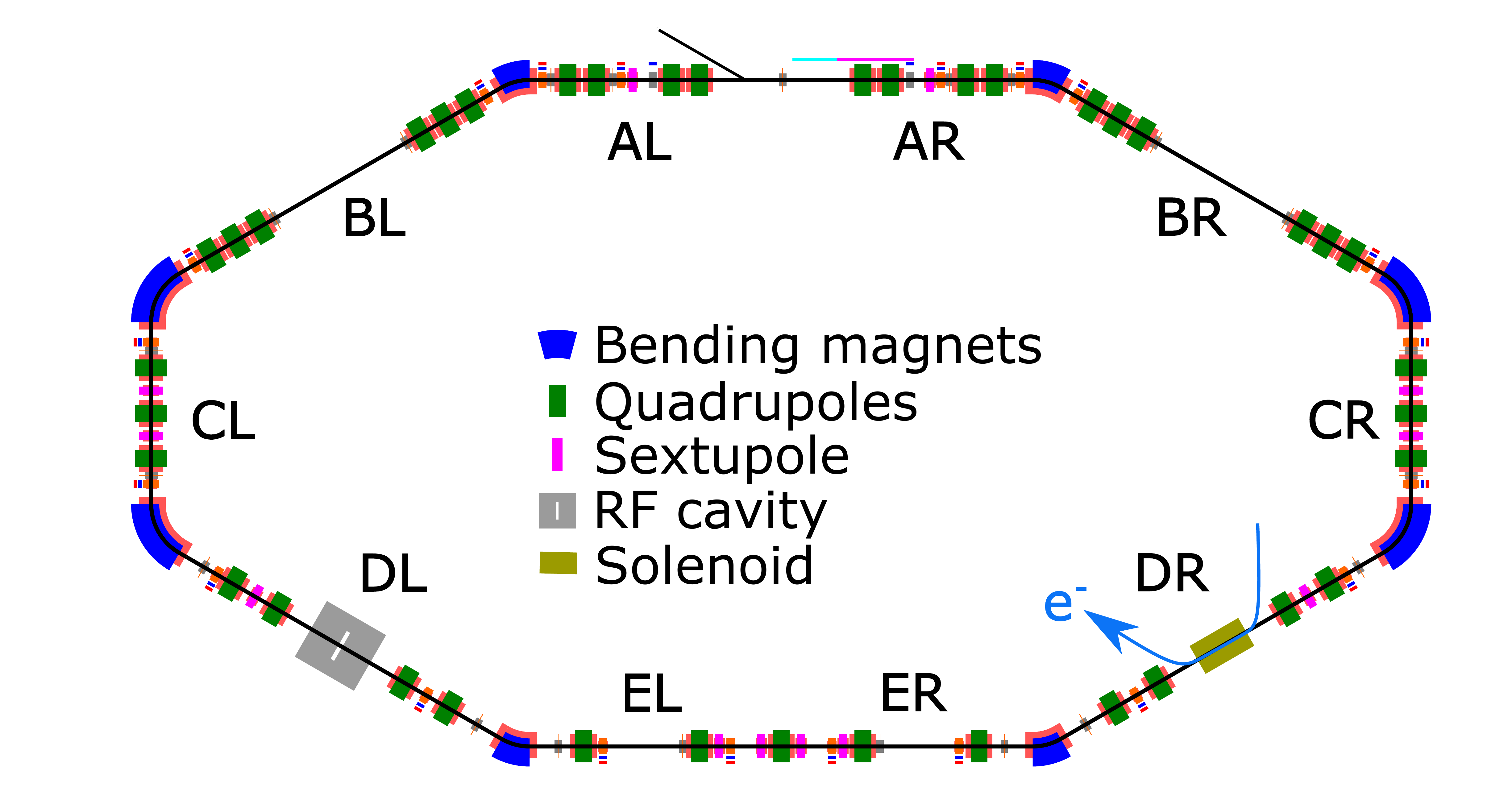}
    \caption{Layout of the Integrable Optics Test Accelerator (IOTA) at Fermilab. The machine is divided into multiple sections named AR (A right), BR (B right) and so on through AL (A left). The blue arrow represents the path of the electrons in the cooler.}
    \label{fig:iota}
\end{figure}

Figure~\ref{fig:iota} shows the layout of IOTA along with the planned location of the electron cooler. Protons with a kinetic energy of 2.5~MeV from the injector (not shown) enter into the ring in the A section, circulate clockwise and co-propagate with electrons in the DR section. Besides the coasting beam mode, an rf cavity operating at the 4\textsuperscript{th} harmonic of the revolution frequency, placed in the DL section is used for bunched beam operation. In addition, non-linear magnets can be placed in the straight sections BL and BR to perform experiments on NIO. While the optics of the ring will be optimized for individual experiments, Table~\ref{tab:protonops} shows some general parameters of operation with electron cooling. At the maximum design current corresponding to a vertical tune shift of -0.5, emittance growth times due to IBS are typically less than 10~seconds, thus limiting beam lifetime and constraining the experiments which can be performed for intense proton beams at IOTA.\footnote{At an effective residual gas pressure of $10^{-10}$~Torr in the IOTA vacuum chamber, the typical emittance growth times due to small angle scattering are in excess of tens of minutes and the large angle scattering lifetime is in the order of 10~hrs. Hence residual gas scattering will not present a significant operational bottleneck for proton operations.} In addition, space-charge forces also create rapid emittance growth and beam loss especially in the first few 100 turns. Consequently electron cooling serves as an important tool to compensate for heating and is valuable for all research with proton beams in IOTA.\\

\begin{table}
    \caption{Typical operation parameters for electron cooling of protons in IOTA. The last three rows list the emittance growth times due to Intra-Beam Scattering (IBS).}
    \label{tab:protonops}
    \centering
    \begin{tabular}{p{\dimexpr 0.4\linewidth-2\tabcolsep}
                    p{\dimexpr 0.18\linewidth-2\tabcolsep}
                    p{\dimexpr 0.18\linewidth-2\tabcolsep}
                    p{\dimexpr 0.13\linewidth-2\tabcolsep}}
        \hline
        Parameter & \multicolumn{2}{c}{Value} & Unit \\
        \hline
        Circumference ($C$) & \multicolumn{2}{c}{39.96} & m \\
        Kinetic energy ($K_b$) & \multicolumn{2}{c}{2.5} & MeV \\
        Revolution time ($\tau_\text{rev}$) & \multicolumn{2}{c}{1.83} & $\mu$s \\
        Emittances ($\epsilon_{x,y}$) & \multicolumn{2}{c}{4.3, 3.0} & $\mu$m \\
        Momentum spread ($\sigma_p/p$) & \multicolumn{2}{c}{$1.32\times10^{-3}$} & \\
        \hline
        & Coasting & Bunched &\\
        \hline
        Number of bunches & - & 4 & \\
        Bunch length ($\sigma_s$) & - & 0.79 & m \\
        Beam current ($I_b$) & 6.25 & 1.24 & mA \\
        Bunch charge ($q_b$) & 11.4 & 0.565 & nC \\
        Tune shifts ($\Delta \nu_{x,y}$) & \multicolumn{2}{c}{-0.38, -0.50} & \\
        $\tau_{\text{IBS},x}$ & 6.40 & 8.69 & s \\
        $\tau_{\text{IBS},y}$ & 4.19 & 5.97 & s \\
        $\tau_{\text{IBS},s}$ & 8.08 & 23.0 & s \\
        \hline
    \end{tabular}
\end{table}

We have chosen a magnetized electron cooler configuration for IOTA with relevant parameters listed in Table~\ref{tab:ecooler}. The electron beam energy and transverse beam size are chosen so that the electrons co-propagate with the protons at the same velocity in the cooler and forms an uniform distribution whose extent overlaps with 2$\sigma$ of the transverse distribution of the protons. The maximum electron current is limited by the space-charge voltage depression at the center of the cooling beam. This is the dominant contribution to the longitudinal temperature $T_{\parallel, \text{SC}}$ of the electrons given by,
\begin{equation}
    T_{\parallel, \text{SC}} = \frac{1}{k_B m_e c^2 \beta^2 \gamma^2}\Big(\frac{1}{2}\frac{I}{4\pi\epsilon_0c\beta}\Big)^2\,,
\end{equation}
where $I$, $\beta$ and $\gamma$ are the electron beam current and relativistic parameters respectively and $k_B$, $m_e$, $c$ and $\epsilon_0$ are physical constants with their usual meaning. The transverse temperature of the electrons is dominated by the cathode temperature $T_\text{cath}$. In the case of a magnetized electron beam, perturbations to the field inside the solenoid manifests itself as an additional effective transverse temperature $T_{\perp,B_\perp}$ given by,\cite{Bridges1978}
\begin{equation}
    T_{\perp,B_\perp} = \frac{m_e a^2 \omega_c^2}{4 k_B} \Big(\frac{B_\perp}{B_\parallel}\Big)^2\,,
\end{equation}
where $a$, $B$, $\omega_c \equiv eB/m_e$ and $B_\perp/B_\parallel$ are the electron beam radius, solenoid field, cyclotron frequency and the field non-uniformity of the cooler solenoid respectively. In the case of magnetized electron cooling, the cyclotron motion of the electrons effectively damps their transverse temperature as seen by the ions, giving rise to a much smaller effective transverse temperature referred to as $T_{\perp,\text{eff}}$ under the \textit{Parkhomchuk} model\cite{Parkhomchuk2000}. Since perturbations to the solenoid field in turn perturbs the cyclotron motion of the electrons, the field flatness of the solenoid must also be constrained and our design specification is such that the $T_{\perp,B_\perp} \lesssim T_{\perp,\text{eff}}$. The initial cooling time with this configuration is about 1~second in all planes for bunched beam.\\

\begin{table}
    \caption{Baseline electron cooler parameters for IOTA.}
    \label{tab:ecooler}
    \centering
    \begin{tabular}{lr}
         \hline
         \multicolumn{2}{c}{Proton parameters} \\
         \hline
         RMS Beam size ($\sigma_{b,x,y}$) & 3.22~mm, 2.71~mm \\
         \hline
         \multicolumn{2}{c}{Electron parameters} \\
         \hline
         Kinetic energy ($K_e$) & 1.36~keV \\
         Current ($I_e$) & 10~mA \\
         Temporal Profile & DC \\
         Transverse Profile & Flat \\
         Radius ($a$) & 6~mm \\
         Source temperature ($T_\text{cath}$) & 1400~K \\
         \hline
         \multicolumn{2}{c}{Main solenoid parameters} \\
         \hline
         Magnetic field ($B_\parallel$) & 0.1~T \\
         Length ($l_\text{cooler}$) & 0.7~m \\
         Field non-uniformity ($B_\perp/B_\parallel$) & $2\times10^{-4}$ \\
         \hline
         \multicolumn{2}{c}{Electron beam temperatures} \\
         \hline
         Longitudinal ($T_{\parallel} \approx T_{\parallel, \text{SC}}$) & 22.3~K \\
         Transverse ($T_\perp \approx T_\text{cath} + T_{\perp,B_\perp}$) & 1407.4~K \\
         Effective ($T_\text{eff} = T_{\perp,\text{eff}} + T_{\parallel}$) & 34.7~K \\
         \hline
         Cooling times ($\tau_\text{park,x,y,s}$) & 1.0~s, 0.9~s, 1.0~s\\
         \hline
    \end{tabular}
    
\end{table}

\begin{figure}
    \centering
    \includegraphics{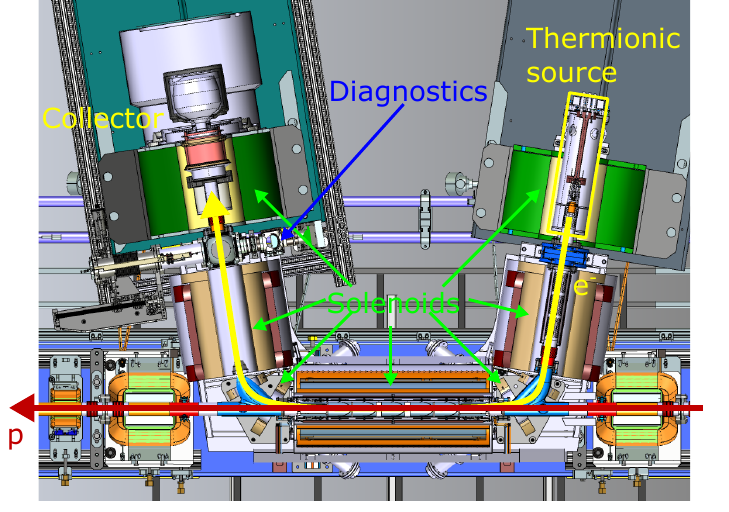}
    \caption{Electron lens setup for IOTA, which will also act as the electron cooler. The total beam power of the magnetized DC electron beam with a flat transverse profile is $\approx 13.6$~W.}
    \label{fig:ecooler}
\end{figure}

\begin{figure*}
    \centering
    \includegraphics[width=6\linewidth/10]{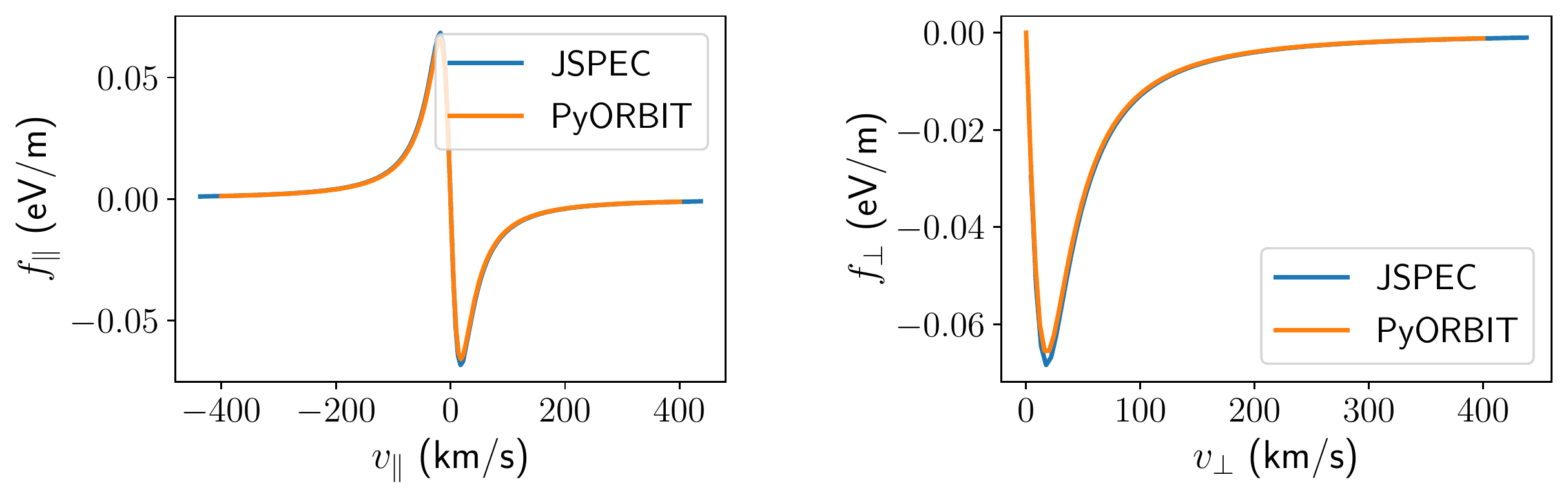}
    \includegraphics[width=3\linewidth/10]{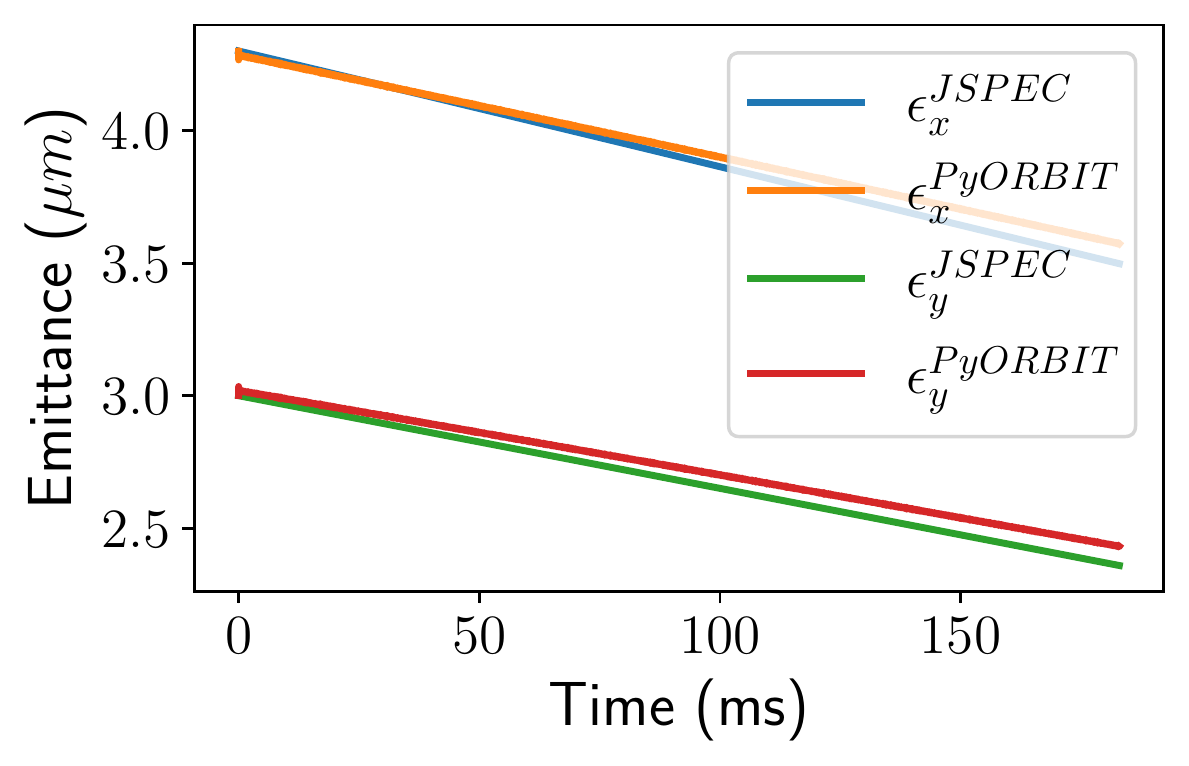}
    \caption{Comparison of the Parkhomchuk cooling force calculated using JSPEC and the \texttt{ecooler} extension for PyORBIT described in this paper. The first two panels show the longitudinal and transverse cooling force as a function of proton velocities and the third panel shows the evolution of transverse emittance as a function of time.}
    \label{fig:benchmark}
\end{figure*}

Figure~\ref{fig:ecooler} depicts a model of the electron cooler setup showing the thermionic source, collector and all the solenoids which keep the electron beam magnetized throughout the transport line. The structure of the thermionic source will be similar to others developed at Fermilab, including for the hollow electron lens at the HL-LHC\cite{Perini2021}. While the strengths of the source and the main solenoid may be adjusted to provide a beam size expansion factor of 2, the baseline design assumes that both the source and the main solenoid are set to 0.1~T. Diagnostics for the electron beam include a Faraday cup to measure current and a scintillating screen to image the transverse profile at the entrance to the collector. Proton beam diagnostics include beam position monitors and a DC current transformer. Additionally, the radiative recombination rate of electrons and protons with the bunched beam parameters listed in Tables~\ref{tab:protonops} and \ref{tab:ecooler} is 6~kHz. This allows for a non-destructive but slow (compared to cooling time) diagnostic of the equilibrium transverse profile of the protons by placing a micro-channel plate detector and associated imaging system\cite{Tranquille2018} at the downstream end of the DR section. The measurement of proton beam lifetime and equilibrium transverse profiles enable the realization of a variety of electron cooling experiments at IOTA.\\

\section{Simulations}

An intense ion beam undergoing electron cooling is subject to focusing from the lattice, cooling forces from the electrons, point-to-point Coulomb scattering with other particles of the beam i.e IBS and also forces due to mean-field contribution from the surrounding charge density i.e. space-charge. All these effects work in concert to limit the equilibrium phase space density which can be reached at the core of the beam. We used the JSPEC code\cite{Zhang2021} to obtain an initial estimate of electron cooling rates which are shown in Table~\ref{tab:ecooler} and also calculate the time evolution of the beam distribution and emittance. Using the Parkhomchuk model of electron cooling and the Martini model of IBS, our simulations indicate the formation of an un-physically dense core\footnote{The IBS models in JSPEC assume a Gaussian distribution of the beam. Hence, large deviations of the cooled beam from the initial Gaussian distribution is the likely cause for the un-physical results and a local model of IBS can alleviate the issue.} which is ruled out by observations.\cite{Nagaitsev95,Steck2000} In order to account for the effects of space-charge forces throughout the accelerator and particles crossing through betatron resonances of the storage ring during the cooling process, we implemented electron cooling as an extension to the particle tracking code PyORBIT.\cite{Shishlo2015}\\

PyORBIT is an extendable particle tracking code implementing symplectic beamline elements and multiple Particle-in-Cell (PIC) space-charge models which allow for the simulation of dynamic space charge effects such as incoherent tune shifts\cite{Titze2020}, emittance growth\cite{Schmidt2016} due to optics function mismatch and heating driven by particles crossing betatron resonances. Space-charge tracking in PyORBIT proceeds as follows, all magnetic lattice elements are divided into smaller segments which are implemented as thick-lens symplectic transfer maps called \emph{nodes} after which thin-lens space-charge kick nodes are placed in between these segments. The beam is tracked through the lattice nodes and at the end of each node, the beam distribution is binned to a Cartesian grid. We use the \texttt{sc2p5d} model of space-charge which assigns transverse kicks to all particles by solving the Poisson's equation in 2D and calculating the total kick the particle receives while travelling through the same length as the last tracked thick-lens segment. In addition to these components we added a new extension to PyORBIT, which inserts a separate \texttt{ecooler} node after every cooler solenoid node. This node calculates a thin-lens kick which accounts for the total momentum change imparted by the cooling force and the static field of the electron beam as the ions travel through a small segment of the cooler. Our novel simulation model thus incorporates XY coupling from the solenoid, static space-charge from the electron beam and dynamic space-charge from the ion beam along with electron cooling.\\

\begin{figure*}
    \centering
    \includegraphics[width=\linewidth]{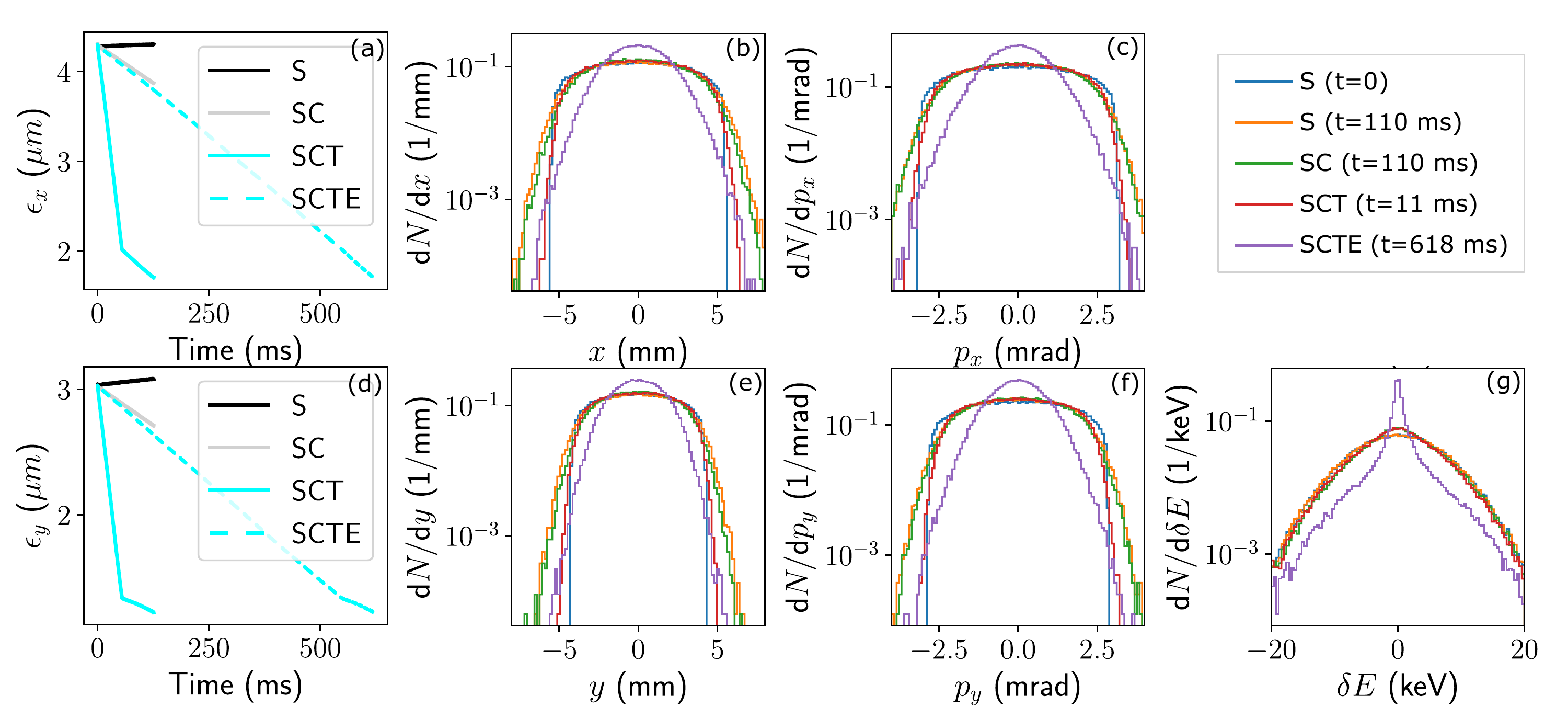}
    \caption{Electron cooling simulation results for IOTA with transverse space charge using the \texttt{eccoler} extension in PyORBIT. Panels (a) and (d) show the evolution of transverse emittance as a function of time for various runs listed in the text. Other panels show snapshots of particle distribution on various axes of phase space.}
    \label{fig:results}
\end{figure*}

The Parkhomchuk model\cite{Parkhomchuk2000} which is used to estimate the cooling force $\vec{F}_\text{cool}$ on each particle in the center-of-momentum (CM) frame of the stored beam is defined as follows:
\begin{equation}
    \begin{split}
        \vec{F}_\text{cool} = - 4 m_e n_e (Z r_e c^2)^2 \frac{\vec{V}}{\Big[ |\vec{V}|^2+\vec{V}_\text{eff}^2\Big]^{3/2}} \\
        \times \ln \bigg[ \frac{\rho_\text{max} + \rho_\text{min} + \rho_L}{\rho_\text{min} + \rho_L} \bigg]\,,
    \end{split}
\end{equation}
where $n_e$, $r_e$, $\rho_L$ and $Z$ are the electron beam density, classical radius, Larmor radius of electrons and the charge of the ions in the storage ring respectively. The cooling force on a single ion is a function of the relative velocity of the ion w.r.t to the electrons given by $\vec{V} \equiv \vec{v}_i - \langle \vec{v}_e \rangle$ and the effective velocity spread of the electrons $V_\text{eff}^2 \equiv \langle v_{e,\parallel}^2 \rangle + \Delta V_{e,\perp}^2$, where $\vec{v}_i$, $\vec{v}_e$, $v_{e,\parallel}$ and $\Delta V_{e,\perp}$ are the CM frame ion velocities, electron velocities, longitudinal component of electron velocities and \emph{effective} transverse component respectively. The effective electron beam temperature listed in Table~\ref{tab:ecooler} is given by $T_\text{eff} \equiv m_e V_\text{eff}^2/k_B$. The range of impact parameters of electron-ion scattering is constrained by $\rho_\text{min} \equiv Z r_e c^2/(|\vec{V}|^2+\vec{V}_\text{eff}^2)$ and $\rho_\text{max} \equiv |\vec{v}_i|/(1/\tau_\text{tof}+\omega_p)$, where $\tau_\text{tof}$ and $\omega_p$ are the time-of-flight of the ion in the cooler and $\omega_p$ is the plasma frequency of the electrons. We assume that $\Delta V_{e,\perp}$ is dominated by the transverse drift velocity in the presence of crossed electric field $E_\rho$ and magnetic field $B_\parallel$ of the cooling beam and the solenoid respectively. Therefore $\Delta V_{e,\perp} \equiv E_\rho (a)/B_\parallel$, where $E_\rho (a)$ is the maximum radial electric field generated by the electron beam. Figure~\ref{fig:benchmark} shows the cooling force estimated using this model in different codes as a function of transverse and longitudinal ion velocities. The third panel in Fig.~\ref{fig:benchmark} compares the evolution of transverse emittance in IOTA as a function of time between JSPEC and our \texttt{ecooler} extension in PyORBIT in the absence of space-charge and IBS. The results show that our implementation of the Parkhomchuk cooling force model agrees with JSPEC in turn-by-turn simulations.\\

The fidelity of the transverse space-charge model \texttt{sc2p5d} used in PyORBIT depends on the transverse grid size and the distance between each space-charge calculation node. In general, the transverse grid size must be much smaller than the Debye length $\lambda_{D,p}$ of the proton beam but also much larger than the mean distance $d_p$ between individual protons. Further the distance between space-charge calculation nodes i.e the time step of the PIC calculation loop $\Delta t_\text{PIC}$ must be much smaller than the plasma period $\tau_p$ of the proton beam. For the proton parameters listed in Table~\ref{tab:protonops}, $\lambda_{D,p} \approx 5$~mm, $d_p \approx 29.1\,\mu$m and $\tau_p \approx 127$~ns. In keeping with these constraints, we choose a $64\times64$ square grid with side length 50~mm which serves as a perfectly conducting boundary for the proton beam and keep the distance between space-charge nodes $< 20$~mm which corresponds to $\Delta t_\text{PIC} \approx 0.91$~ns. As the beam cools and the proton beam density increases, $\lambda_{D,p}$, $d_p$ and $\tau_p$ all reduce in value. However in the simulations reported here, we use fixed discretization parameters. For a coasting beam at the design current of 6.25~mA, the space-charge forces dominate the dynamics over electron cooling with the baseline parameters listed in Table~\ref{tab:ecooler} and the simulations predict long term emittance growth. Simulations starting with a tune shift of $\Delta \nu_y = -0.1$ corresponding to a coasting beam current of 1.25~mA show interesting dynamics and some preliminary results are depicted in Fig.~\ref{fig:results}.\\

The electron cooling simulations reported here are named as follows: S represents a simulation with space-charge only without electron cooling, SC represents a run with space-charge and baseline parameters of electron cooling and SCT represents a run with space-charge and electron-cooling where the cooling force is scaled by a factor of 10 for the first 30000 turns ($\approx 54$~ms) and then switched back to baseline values for the rest of the simulation. The transverse emittance grows for run S while it decays for the run SC, as seen in panels (a) and (d) of Fig.~\ref{fig:results}, indicating that electron cooling dominates the dynamics at the initial stage of the simulation. Looking at the 1D histograms of particle distribution in $x$, $p_x$, $y$ and $p_y$ at $t=110$~ms in panels (b), (c), (e) and (f) respectively indicate that space-charge forces lead to the diffusion of particles from the core of the beam into the periphery as seen in the orange (run S) and green (run SC) traces which are very similar to each other and in general broader than the injected beam (K-V distribution in transverse) shown by the blue traces. In contrast, the energy distribution of particles in panel (g) shows a clear evidence of cooling for run SC (green) as evidenced by the increase of particles at low energy deviations $\delta E$ when compared to the histogram for S (orange) at the same time. Even though the runs S and SC yield some insight into this initial stage of cooling when the space-charge forces are weak, longer term simulations are required to probe the intense space-charge regime.\\

In order to speed up the simulation in the initial weak space-charge regime, we introduce the run SCT where the cooling force is scaled by a factor of 10 during an initial period where space-charge forces are much smaller compared to the linear focusing provided by the lattice. The transverse emittance decays much faster for the run SCT compared to the run SC, but when the time axis is scaled by a factor of 10 (labelled SCTE in panels (a) and (d) of Fig.~\ref{fig:results}), the emittance decay closely follows SC. Comparing the distribution of particles in phase-space at $t=110$~ms for SC (green) and $t=11$~ms for SCT (red), we see very similar distributions in energy as shown by panel (g) and good agreement up to intermediate values of the transverse coordinates $x$, $p_x$, $y$ and $p_y$ in panels (b), (c), (e) and (f) respectively. Hence by suitably scaling the time for the run SCT, we can estimate the phase space distribution of the beam at $t=618$~ms as shown in purple in Fig.~\ref{fig:results}. The results indicate reduction in transverse emittance by almost a factor of 2 in both planes and also significant reduction in energy spread.\\

\section{Conclusion and Outlook}
While electron cooling provides a mechanism to cool ions of low to intermediate kinetic energies, the maximum intensity of the ion beam attained through this method is limited by Intra Beam Scattering (IBS) and space-charge forces acting on the ion beam. A maximum incoherent space-charge tune shift of 0.1-0.2 has been achieved in practice. We are developing an electron cooling test bed for 2.5~MeV protons at the Integrable Optics Test Accelerator (IOTA) at Fermilab which is a 40~m storage ring operating up to a design current corresponding to a transverse space-charge tune shift of $\Delta \nu_y = -0.5$. We have developed the baseline design of a magnetized electron cooler which will enable the study of electron cooling for ion beams with intense space-charge, study the interplay between space-charge and instabilities and use cooling as a method of phase space control for Non-linear Integrable Optics.\\

With the baseline parameters, the cooling times are of the order of one second in all planes which is sufficient to compensate for emittance growth due to IBS at the planned proton beam parameters. To study the influence of intense space-charge with electron cooling, we developed an extension to the particle tracking code PyORBIT which uses the Parkhomchuk model of the cooling force, a static model of the electron beam space-charge and a transverse dynamic space-charge model of the proton beam. While we verified the cooling force with the simulation code JSPEC, efforts to validate long-term space-charge tracking in PyORBIT are still ongoing. Numerical simulations using our tool show that on starting with a coasting beam of 1.25~mA in IOTA corresponding to a tune shift of $\Delta \nu_y = -0.1$, the transverse emittances reduce by a factor of 2 in $618$~ms. However the effect of space-charge is still weak and longer simulations covering multiple cooling times is required to adequately study the intense space-charge regime. This effort represents the first time electron cooling simulations have incorporated turn-by-turn space-charge tracking with all lattice elements enabling the analysis of non-linear dynamics of the particles, crossing of betatron resonances, etc.\\

Future simulation work will involve using an amplitude dependent model of IBS with electron cooling to compare with space-charge simulation results, studying the effect of a semi-hollow electron beam distribution on the cooling process and developing the thermionic electron source to generate the required beam. Based on these calculations, we will develop specific plans for experimental study in IOTA.

\section*{Acknowledgement}
We would like to thank Alexander Valishev, Alexander Romanov and Valeri Lebedev for discussions on proton operations in IOTA and helping develop the experimental parameters. We also want to acknowledge Alexey Burov and Alexei Fedotov for recognizing the need for a local model of IBS to prevent over-cooling in our JSPEC simulations. This manuscript has
been authored by Fermi Research Alliance, LLC under Contract No.~DE-AC02-07CH11359 with the U.S.\ Department of Energy, Office of
Science, Office of High Energy Physics. This research is also
supported by the University of Chicago.


\begin{thebibliography}{99}
    
    \bibitem{Spiller2020}
    P.~Spiller \emph{et al.}, \textquotedblleft{The FAIR Heavy Ion Synchrotron SIS100}\textquotedblright, JINST \textbf{15}, no. 12, T12013, 2020
    
    \bibitem{Yang2013}
    J.~C.~Yang \emph{et al.}, \textquotedblleft{High Intensity heavy ion Accelerator Facility (HIAF) in China}\textquotedblright, Nucl. Instrum. Meth. B \textbf{317}, 263-265, 2021.
    
    \bibitem{Kekelidze2017}
    V.~D.~Kekelidze., \textquotedblleft{NICA project at JINR: status and prospects}\textquotedblright, JINST \textbf{12}, no. 06, C06012, 2017.
    
    \bibitem{Willeke2021}
    F.~Willeke \emph{et al.}, \textquotedblleft{Electron Ion Collider Conceptual Design Report 2021}\textquotedblright, United States, 2021. \url{https://www.bnl.gov/ec/files/EIC_CDR_Final.pdf}
    
    \bibitem{Bartmann2018}
    W.~Bartmann \emph{et al.} [ELENA and AD],\textquotedblleft{The ELENA facility}\textquotedblright, Phil. Trans. Roy. Soc. Lond. A \textbf{376}, no.2116, 20170266, 2018
    
    \bibitem{Nagaitsev2006}
    S.~Nagaitsev, \emph{et al.}, \textquotedblleft{Experimental demonstration of relativistic electron cooling}\textquotedblright, Phys. Rev. Lett. \textbf{96}, 044801, 2006
    
    \bibitem{Parkhomchuk2001}
    V.~Parkhomchuk, I.~Ben-Zvi,, \textquotedblleft{Electron Cooling for RHIC}\textquotedblright, BNL C-A/AP/47, 2001. \url{https://www.bnl.gov/isd/documents/79867.pdf}
    
    \bibitem{Nagaitsev95}
    S. Nagaitsev \emph{et al.}, \textquotedblleft{Space Charge Effects and Intensity Limits of Electron-Cooled Bunched Beams}\textquotedblright, in \emph{Proc. 16th Particle Accelerator Conf. (PAC’95)}, Dallas, TX, USA, May 1995, paper RAP22, pp. 2937--2939.
    
    \bibitem{Steck2000}
    M.~Steck, \emph{et al.}, \textquotedblleft{Beam accumulation with the SIS electron cooler}\textquotedblright, Nucl. Instrum. Meth. A \textbf{441}, 175-182, 2000.
    
    \bibitem{Antipov2016}
    S.~Antipov, \emph{et al.}, \textquotedblleft{IOTA (Integrable Optics Test Accelerator): Facility and Experimental Beam Physics Program}\textquotedblright, JINST \textbf{12}, no.03, T03002, 2017.
    
    \bibitem{Valishev2021}
    A. Valishev, \emph{et al.}, \textquotedblleft{First Results of the IOTA Ring Research at Fermilab}\textquotedblright, in \emph{Proc. IPAC'21}, Campinas, SP, Brazil, May 2021, pp. 19--24.
    
    \bibitem{Stancari2021}
    G.~Stancari, \emph{et al.}, \textquotedblleft{Beam physics research with the IOTA electron lens}\textquotedblright, JINST \textbf{16}, no.05, P05002, 2021.
    
    \bibitem{Burov2018}
    A.~Burov, \textquotedblleft{Convective Instabilities of Bunched Beams with Space Charge}\textquotedblright, Phys. Rev. Accel. Beams \textbf{22}, no.3, 034202, 2019.
    
    \bibitem{Bridges1978}
    J.~Bridges, \emph{et al.}, \textquotedblleft{Fermilab Electron Cooling Experiment: Design Report}\textquotedblright, FERMILAB-DESIGN-1978-01, 1978.
    \url{https://lss.fnal.gov/archive/design/fermilab-design-1978-01.pdf}
    
    \bibitem{Parkhomchuk2000}
    V.~V.~Parkhomchuk, \textquotedblleft{New insights in the theory of electron cooling}\textquotedblright, Nucl. Instrum. Meth. A \textbf{441}, 9-17, 2000.
    
    \bibitem{Perini2021}
    D.~Perini, A.~Kolhemainen, A.~Rossi, S.~Sadovich and G.~Stancari, \textquotedblleft{Design of high-performance guns for the HL-LHC HEL}\textquotedblright, JINST \textbf{16}, no.03, T03010, 2021.
    
    \bibitem{Tranquille2018}
    G. Tranquille \emph{et al.}, \textquotedblleft{Commissioning the ELENA Beam Diagnostics Systems at CERN}\textquotedblright, in \emph{Proc. 9th Int. Particle Accelerator Conf. (IPAC’18)}, Vancouver, Canada, Apr.-May 2018, pp. 2043--2046.
    
    \bibitem{Zhang2021}
    H. Zhang, S. V. Benson, M. W. Bruker, and Y. Zhang, \textquotedblleft{JSPEC - A Simulation Program for IBS and Electron Cooling}\textquotedblright, in \emph{Proc. 12th Int. Particle Accelerator Conf. (IPAC’21)}, Campinas, Brazil, May 2021, pp. 49--52.
    
    \bibitem{Shishlo2015}
    A.~Shishlo, S.~Cousineau, J.~Holmes, T.~Gorlov, \textquotedblleft{The Particle Accelerator Simulation Code PyORBIT}\textquotedblright, Procedia Computer Science, \textbf{51}, 2015, pp. 1272--1281.
    
    \bibitem{Titze2020}
    M.~Titze, \textquotedblleft{Space Charge Modeling at the Integer Resonance for the CERN PS and SPS}\textquotedblright, PhD thesis, Humboldt-Universität zu Berlin, Germany, 2020. \url{https://doi.org/10.18452/21423}
    
    \bibitem{Schmidt2016}
    F. Schmidt \emph{et al.}, \textquotedblleft{Code Bench-Marking for Long-Term Tracking and Adaptive Algorithms}\textquotedblright, in \emph{Proc. 57th ICFA Advanced Beam Dynamics Workshop on High-Intensity and High-Brightness Hadron Beams (HB’16)}, Malmö, Sweden, Jul. 2016, pp. 357--361.
\end{thebibliography}
\end{document}